\documentclass[11pt,prd,twocolumn,nofootinbib,reprint,superscriptaddress,longbibliography,colorlinks=true,citecolor=blue]{revtex4-1}
\usepackage{amsmath,amssymb, amsthm, graphicx, epsfig, fancyhdr,epsfig,multirow}
\usepackage[normalem]{ulem}
\usepackage[dvipsnames]{xcolor}
\usepackage{tikz-feynman}
\tikzfeynmanset{compat=1.1.0}

\usepackage{tikzsymbols}
\usepackage{natbib}
\usepackage{xcolor}
\usepackage{hyperref}

\makeatletter
\def\alt{\mathrel{\mathpalette\gl@align<}}
\def\agt{\mathrel{\mathpalette\gl@align>}}
\def\gl@align#1#2{\lower.6ex\vbox{\baselineskip\z@skip\lineskip\z@
\ialign{$\m@th#1\hfil##\hfil$\crcr#2\crcr\sim\crcr}}} \makeatother

\def\beq{\begin{equation}}
\def\eeq{\end{equation}}
\def\bea{\begin{eqnarray}}
\def\eea{\end{eqnarray}}

\pdfoutput=1

\begin{document}

\pagestyle{plain}

\title{Exclusion limits on Dark Matter-Neutrino Scattering Cross-section}

\author{Diptimoy Ghosh}
\email{diptimoy.ghosh@iiserpune.ac.in}
\affiliation{Department of Physics, Indian Institute of Science Education and Research Pune, Dr Homi Bhaba Road, NCL Colony, Pashan, Pune, Maharashtra - 411008, India}

\author{Atanu Guha}
\email{atanu@cnu.ac.kr}
\affiliation{Department of Physics, Indian Institute of Science Education and Research Pune, Dr Homi Bhaba Road, NCL Colony, Pashan, Pune, Maharashtra - 411008, India}
\affiliation{Department of Physics, Chungnam National University, 99, Daehak-ro, Yuseong-gu, Daejeon-34134, South Korea}

\author{Divya Sachdeva}
\email{dsachdeva@lpthe.jussieu.fr}
\affiliation{Department of Physics, Indian Institute of Science Education and Research Pune, Dr Homi Bhaba Road, NCL Colony, Pashan, Pune, Maharashtra - 411008, India}
\affiliation{Laboratoire de Physique Théorique et Hautes Energies (LPTHE),
UMR 7589 CNRS \& Sorbonne Université, 4 Place Jussieu, F-75252, Paris, France}

\begin{abstract}
We derive new constraints on combination of dark matter - electron cross-section ($\sigma_{\chi e}$) and dark matter - neutrino cross-section ($\sigma_{\chi \nu}$) utilising
the gain in kinetic energy of the dark matter (DM) particles due to scattering with the cosmic ray electrons and the diffuse supernova neutrino background (DSNB). Since the flux
of the DSNB neutrinos is comparable to the CR electron flux in the energy range $\sim 1\,{\rm MeV}  - 50 \,{\rm MeV}$, scattering with the
DSNB neutrinos can also boost low-mass DM significantly in addition to the boost due to  interaction with the cosmic ray electrons. We use the XENON1T as well as the
Super-Kamiokande data to derive bounds on $\sigma_{\chi e}$ and  $\sigma_{\chi \nu}$. While our bounds for $\sigma_{\chi e}$ are comparable with those in the literature, we show that the Super-Kamiokande experiment provides the  strongest constraint on $\sigma_{\chi \nu}$ for DM masses below a few MeV.

\end{abstract}
\maketitle

\section{Introduction}
The presence of dark matter (DM) in the universe and its dominance over luminous matter is well-established. DM is known to be non-relativistic and very weakly interacting
with the Standard Model (SM) particles. The existence of DM has been inferred only via their gravitational interactions, and both the DM mass and
their exact interaction strengths with the SM particles are unknown (see \cite{Bauer:2017qwy,Bertone:2004pz,Lisanti:2016jxe,Profumo:2013yn} for reviews).

The DM-proton and DM-electron interaction strengths have been constrained severely by various direct detection (DD)
experiments~\cite{XENON:2020iwh,XENON:2020rca,Crisler:2018gci,CRESST:2015txj,DarkSide:2018ppu}. Their basic working principle is to measure the recoil
energy of electron/nucleus once the incoming DM particles scatter of them. But one major limitation of this method is that these DD experiments lose
sensitivity rapidly for low mass DM (below DM mass of $\sim$ 5 GeV for DM-proton cross-section~\cite{XENON:2015gkh}, and below DM mass of $\sim$ 5 MeV for DM-electron cross-section~\cite{Essig:2012yx})
since very light DM particles cannot produce enough recoil to be detected.
The most stringent constraints on DM-electron scattering cross-section comes from the SENSEI experiment: $\sigma_{\chi\,e}\lesssim \, 10^{-34} \,\rm{cm^2}$ for $m_\chi > 5\,{\rm MeV}$~\cite{SENSEI:2019ibb}. 

In the past few decades many other novel detection strategies have been developed: employing the energy deposition in white dwarfs or neutron stars\cite{Baryakhtar:2017dbj,Dasgupta:2019juq,Bose:2021yhz,Guha:2021njn,Sen:2021wev,Acevedo:2020gro,Leane:2020wob,Garani:2020wge,Joglekar:2020liw,Joglekar:2019vzy},
from Big Bang Nucleosynthesis\cite{Green:2017ybv,Baumann:2016wac,Krnjaic:2019dzc}, DM production in astrophysical objects~\cite{McKeen:2020vpf}, cooling of stars and supernovae\cite{Raffelt:2006cw,Raffelt:1999tx,Kadota:2014mea,Dreiner:2013mua,Guha:2018mli}, colliders searches etc. \cite{Bai:2011wy,Cohen:2015toa}.
Cosmological observations have also been used to constrain interactions of low mass DM~\cite{Ali-Haimoud:2021lka,Nguyen:2021cnb}. For example, CMB spectral distortion excludes DM-electron
scattering cross-section above $\sigma_{\chi\,e}  \sim 10^{-28}{\rm cm^2}$ for $m_\chi \lesssim 0.1\,{\rm MeV}$~\cite{Ali-Haimoud:2021lka} if the cross section is momentum independent.

The DM-neutrino cross-section ($\sigma_{\chi\nu}$), which is the focus of our study, cannot be independently constrained by the DD experiments since matter is not made of neutrinos. 
However, as we will discuss later, the DD experiments can be used to constrain a function of $\sigma_{\chi\nu}$ and $\sigma_{\chi e}$ (for more details refer to Sec.~\ref{Sec:rate}). 
Independent bound on $\sigma_{\chi\nu}$ does however exist from Planck and large scale structure experiments: $\sigma_{\chi\nu}\lesssim \, 10^{-33} (m_\chi/{\rm GeV)\,cm^2}$
if the cross section is momentum independent~\cite{Wilkinson:2014ksa}. {Such constraints are especially important for those models where DM interact with leptons and dominant interaction is with neutrinos~\cite{Brune:2018sab,Blennow:2019fhy}.}

\begin{figure}
\centering
\includegraphics[width=0.49\textwidth]{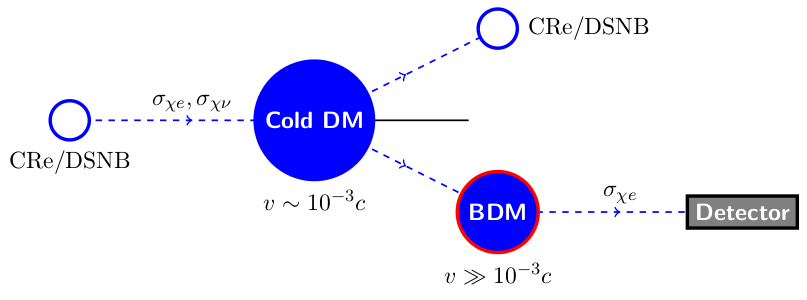}
\caption{Schematic diagram for production of boosted dark matter (BDM) due to scattering with CR electrons (CRe) and DSNB neutrinos, and their subsequent detection at the detectors.}
\protect\label{Fig:schematic}
\end{figure}

In order to recover the sensitivity of the DD experiments for low mass DM, the idea of Boosted Dark Matter (BDM) has been explored in recent years.
In this scenario, the DM is boosted to higher velocity due to scattering with various cosmic ray components (see  Fig.~\ref{Fig:schematic} for a schematic diagram)
This idea has been explored using CR protons~\cite{Bringmann:2018cvk,Cappiello:2018hsu,Cappiello:2019qsw}, helium nuclei~\cite{Bringmann:2018cvk}, 
cosmic electrons~\cite{Ema:2018bih,Cappiello:2018hsu,Cappiello:2019qsw,Dent:2020syp,Jho:2020sku,Bramante:2021dyx}, and neutrinos from various astrophysical sources~\cite{Farzan:2014gza,Arguelles:2017atb,Yin:2018yjn,Jho:2021rmn,Das:2021lcr}. 

Our Universe is abundant in MeV energy neutrinos emerging from massive stars going supernova, right from the epoch of first star formation.
Similarly, our galaxy also has energetic electron flux. We consider  that the DM in the Milky Way halo experiences scattering with the DSNB neutrinos as well as the cosmic electrons, and gets boosted to velocities $v\gg\,10^{-3}\,{\rm c}$. Such upscattered low-mass DM can leave interesting signatures in low-energy recoil experiments like XENON1T (where detectable electron recoil energies are of the order of few keV because it is based on the scintillation process), as well as in high-energy recoil experiments like Super-Kamiokande (where detectable electron recoil energies are of the order of few MeV because Super-K relies on Cherenkov radiation). 

In earlier studies, CR electrons and DSNB were separately considered as the particles transferring kinetic energy to the DM. However, as shown in Fig.~\ref{Fig:cr_flux}, the fluxes for
CR electrons and DSNB neutrinos are comparable in the energy range $1\,{\rm MeV} \lesssim \,{\rm T_i}\, \lesssim  50\,{\rm MeV}$ motivating us to include both these contributions
to the DM boost. Note that, while studying $\sigma_{\chi e}$, $\sigma_{\chi\nu}$ can in principle be set to zero but the vice versa cannot be done since, for the detection of
the DM signal, the existence of $\sigma_{\chi e}$ is crucial. And once $\sigma_{\chi e}$ is non-zero, it will also contribute to boost the DM and must be taken into account in a consistent analysis.
In this work, we have thus taken into account contributions to the DM boost due to the CR electrons as well as the DSNB neutrinos, and used the 
 XENON1T~\cite{XENON:2020rca} and Super-Kamiokande~\cite{Super-Kamiokande:2011lwo} data to provide exclusion limit on the DM-neutrino and DM-electron interactions. To our knowledge, we are the first to use the Super-Kamiokande data to constrain the  DM-neutrino cross-section, and we show that it provides the strongest direction detection bound in the DM mass range below $\sim \, 10$ MeV. 
 
 The paper is organized as follows. In Sec. II, we discuss the DSNB and cosmic ray electron flux as a function of energy. In Sec. III, we calculate the flux of boosted dark matter which is used to calculate rate of scattering (Sec. IV) in different detectors. This rate is then compared and analyzed with the observations of Xenon1T and Super Kamiokande using $\chi^2$ statistics in section V. In Sec. VI, we discuss our results and conclude.

\section{Cosmic ray electron and neutrino flux}
\label{Sec:neutrino_flux}
CR electron flux can be described by certain parameterization of the local interstellar spectrum~\cite{Boschini:2018zdv} given as
\begin{widetext}
\begin{align}
  F(T_e) = \begin{cases}
     \mbox{\Large\( \frac{1.799 \times 10^{44}~ T_e^{-12.061}}{1+ 2.762 \times 10^{36}~ T_e^{-9.269} + 3.853 \times 10^{40}~ T_e^{-10.697}} \)} & \text{if $T_e < 6880$ MeV} \\
   \mbox{\tiny\( \)} & \mbox{\tiny\( \)} \\
      3.259 \times 10^{10}~T_e^{-3.505} + 3.204 \times 10^{5}~T_e^{-2.620} & \text{if $T_e \geqslant 6880$ MeV}
    \end{cases}
\label{Eq:CRe_flux_parameterization}
\end{align}
\end{widetext}
where the unit of $F(T_e)$ is given in $\rm{\left(m^2~s~sr~MeV \right)^{-1}}$ and the kinetic energy ($T_e$) of the CR electrons is in MeV. The above parameterization is in accordance with Fermi-LAT~\cite{Fermi-LAT:2011baq,Fermi-LAT:2009yfs,Fermi-LAT:2010fit,Fermi-LAT:2017bpc}, AMS-02~\cite{AMS:2014gdf}, PAMELA~\cite{PAMELA:2011bbe,CALET:2017uxd}, and Voyager~\cite{Cummings:2016pdr,Stone:2013} data.  
 
On the other hand, the DSNB flux can be calculated from the knowledge of the rate of core-collapse supernovae ($R_{CCSN}$)~\cite{Beacom:2010kk, Horiuchi:2008jz}
 
 \bea
\Bigl. \Phi_{\nu}(E_\nu)\Bigr\vert_{DSNB} = \int_0^{z_{max}} \frac{dz}{H(z)} R_{CCSN}(z) F_\nu(E'_\nu)
 \eea
 
 The $R_{CCSN}$ is a function of star formation rate and has been calculated following~\cite{Lunardini:2010ab}. The observable effective spectra of the neutrinos emitted from supernovae, is assumed to be of Fermi-Dirac form and approximately given for each flavour as~\cite{Beacom:2010kk,Lunardini:2010ab}
 \bea
 F_\nu(E_\nu) = \frac{E_\nu^{tot}}{6} \frac{120}{7 \pi^4} \frac{E_\nu^2}{T_\nu^4} \frac{1}{\exp(\frac{E_\nu}{T_\nu}) + 1}
 \eea
 
 where $ E_\nu^{tot} \approx 5 \times 10^{52}~\rm{erg}$ is the total energy released in the supernova explosion in the form of neutrinos. $T_\nu$ is the temperature of the neutrinos.  Supernova simulations and Super-Kamiokande observations suggest the estimates of the temperatures to be $T_\nu = 4~\rm{MeV}$ for $\nu_e$, $T_\nu = 5~\rm{MeV}$ for $\bar{\nu}_e$ and $T_\nu = 8~\rm{MeV}$ for $\nu_\mu, \bar{\nu}_\mu, \nu_\tau, \bar{\nu}_\tau$~\cite{Lunardini:2010ab,Super-Kamiokande:2013ufi}. $E'_\nu = E_\nu (1+z)$ is the observed energy at earth due to redshift $z$. The maximum redshift is generally taken to be $z_{max}=5$. The Hubble function is given by $H(z)=H_0 \sqrt{\Omega_\Lambda + \Omega_m (1+z)^3}$, where, $H_0 = 67~\rm{km s^{-1} Mpc^{-1}}$. $\Omega_\Lambda = 0.7$ and $\Omega_m = 0.3$ are the vacuum and matter contribution to the energy density of the Universe, respectively~\cite{Planck:2018vyg}.

\begin{figure}
\centering
\includegraphics[width=0.49\textwidth]{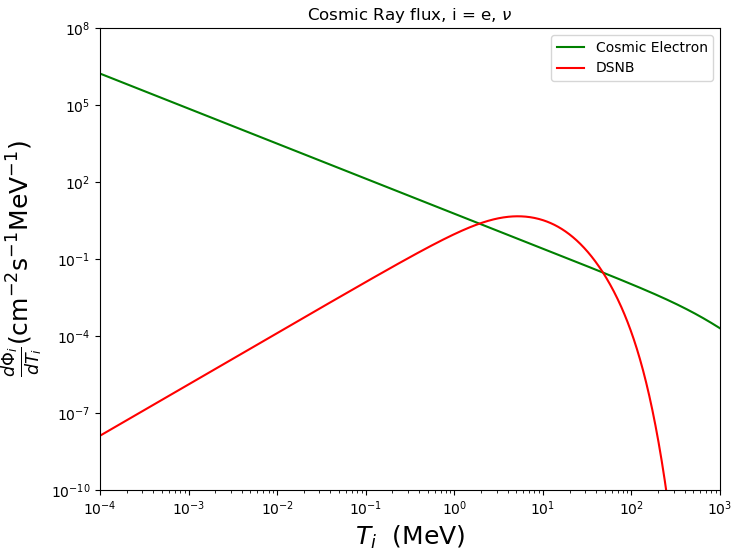}
\caption{Fluxes of cosmic ray electrons and the diffuse supernova neutrino background (summing over all flavours of neutrinos).}
\protect\label{Fig:cr_flux}
\end{figure}

\begin{figure*}
\centering
\includegraphics[width=0.49\textwidth]{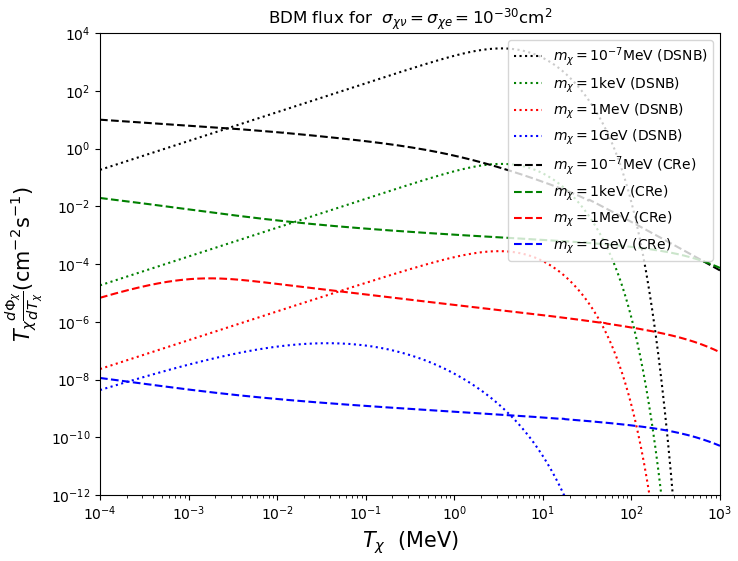}\hfill
\includegraphics[width=0.49\textwidth]{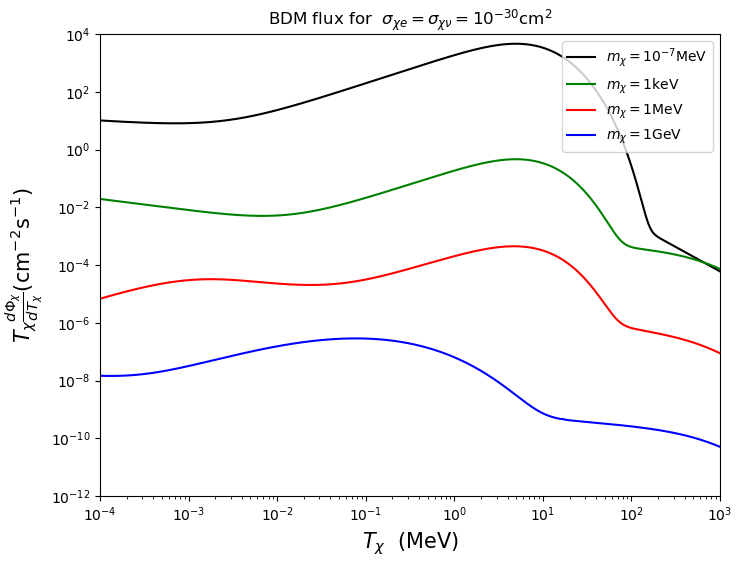}
\caption{CR electrons, DSNB induced BDM flux for $\sigma_{\chi e} = \sigma_{\chi \nu} = 10^{-30}~\rm{cm^2}$. The left panel shows the BDM flux due to the individual boost by CR electrons (dashed lines), DSNB neutrinos (dotted lines). The right panel shows the BDM flux due to the combined effect.}
\protect\label{Fig:bdm_flux}
\end{figure*}

Fig.~\ref{Fig:cr_flux} shows the cosmic ray electron flux along with the flux of neutrinos coming from supernovae. As can be observed from this figure, for $1\,{\rm MeV}<\,{\rm T_i}\,<50\,{\rm MeV}$, both the fluxes are of similar order and this range is crucial for boosting DM with mass below MeV. This fact inspire us to consider both the fluxes to derive bounds on $\sigma_{\chi\nu}$ and $\sigma_{\chi e}$ for light DM case considered here.


\section{Boosted Dark matter flux}
\label{Sec:bdm_flux}

 Cold dark matter (DM) particles get boosted after getting hit by cosmic ray electrons and DSNB. { We assume that the DM-electrons or DM-neutrinos scattering cross-section is constant as a function of center of mass energy.} We calculated the flux of the boosted DM using the cosmic ray electron and neutrino fluxes discussed at Sec.~\ref{Sec:neutrino_flux} following~\cite{Bringmann:2018cvk,Ema:2018bih,Cappiello:2018hsu}.
 
For $i$-DM scattering ($i=e,\nu$), the energy transfer to the cold DM by the CR electron/DSNB is given by
    \bea
  T_{\chi} &=&  T_{\chi}^{max}  \left( \frac{ 1 - \cos \theta }{2}\right)\nonumber \\ T_{\chi}^{max} &=& \frac{\left(T_i\right)^2+ 2 T_i m_i  }{T_i+ \left(m_i +m_{\chi} \right)^2/\left(2 m_{\chi} \right)} 
  \label{Eq:T-chi-max} 
  \eea
 where $\theta$ is the scattering angle at the centre of momentum frame.
 
 Solving Eq.~\ref{Eq:T-chi-max} we get the minimum required energy of the electrons/neutrinos to produce a certain amount of kinetic energy of the boosted DM
  
  \bea
  T_i^{min} = \left( \frac{T_{\chi}}{2} -m_i \right) \left[ 1 \pm \sqrt{ 1 +  \frac{2 T_{\chi}}{ m_{\chi}} \frac{\left(m_i +m_{\chi}\right)^2}{\left( 2m_i - T_{\chi} \right)^2}} \right] 
  \label{Eq:T-nu-min}
  \eea
  
  The $+$ and $-$ sign in Eq.~(\ref{Eq:T-nu-min}) are applicable for $T_\chi > 2 m_i$ and $T_\chi < 2 m_i$, respectively. 
  
  For the elastic scattering cross-section $\sigma_{\chi i}$, the collision rate of $i$-$\chi$ per unit volume, having kinetic energy of the particle $i$ in the range $\left[ T_i, T_i + dT_i \right]$, is given by
 \bea
 \frac{d\Gamma}{dV} &=& \sigma_{\chi i} \times n_{\chi} \times d\Phi_i \nonumber \\
 d\Gamma &=& \sigma_{\chi i} \times \frac{\rho_\chi}{m_\chi} \times \frac{d\Phi_i}{dT_i}~dT_i~dV
 \eea
 
  Where, $\frac{d\Phi_i}{dT_i}$ is the Local Interstellar Spectrum for the particle $i$. The solid angle subtended by a small facet having flat surface area $ds$, orientation $\hat{n}$ and distance $d$ from the viewer is
 \bea
 d\Omega = 4\pi \left( \frac{ds}{4\pi d^2}\right) \left(\hat{d}\cdot \hat{n} \right)
 \eea 
 If $d$ is very large, $\hat{d} \cdot \hat{n} \approx 1$.
 
 We consider that the incoming electrons/neutrinos with kinetic energy $\left[ T_i, T_i + dT_i \right]$  are contained within $dV$. Let $dl$  be the length at which $dV$ is spread along $\hat{d}\approx\hat{n}$ and $ds \hat{n}$ be the area vector of the flat surface. In that case, $ dV \approx dl \cdot ds$ and $4 \pi d^2$ is the area of the sphere of radius $d$ around the DM particles.
 
 Electrons/neutrinos are going out from $dV$ to all direction. But we are interested only those who are directed towards the DM. Therefore, $i$ induced DM flux is given by
 \bea
 \frac{d\Phi_\chi}{dT_i} &=& \int \sigma_{\chi i} \cdot  \frac{\rho_\chi}{m_\chi} \cdot \frac{d\Phi_i}{dT_i} \cdot dl \cdot \frac{ds}{4 \pi d^2} \nonumber \\
 &=& \int_{\rm{Line~of~sight}} \sigma_{\chi i} \cdot  \frac{\rho_\chi}{m_\chi} \cdot \frac{d\Phi_i}{dT_i} \cdot dl \cdot \frac{d\Omega}{4 \pi}
 \eea
 
 Now we have to integrate over all possible line segment $dl$ along the line of sight. Let $D_{eff}$ be the effective distance out to which all the electrons/neutrinos have to be taken into account. So, for homogeneous and isotropic distribution of DM and $i$ 
 \bea
 \frac{d\Phi_{\chi}}{dT_i} = \sigma_{\chi i} \times \frac{\rho_\chi}{m_\chi} \times \frac{d\Phi_i}{dT_i} \cdot D_{eff}
 \eea
 
 In our present work, we consider the effective distance for the DSNB to be $D_{eff}=10~\rm{kpc}$ and for the incoming cosmic ray electrons $D_{eff}=1~\rm{kpc}$.
 As an independent variable, if $T_\chi$ is having flat distribution (i.e., $T_\chi$ can take any value in the range $\left[ 0,T_\chi^{max} \right]$ with equal probability), we can write, 
 \bea
 \frac{d\Phi_\chi}{dT_\chi} =  \frac{\Phi_\chi}{T_\chi^{max}}~ \Theta(T_\chi^{max} - T_\chi)
 \eea
 
 But $\Phi_\chi$ and $T_\chi^{max}$ both are functions of $T_i$, and  in that case $\Phi_\chi$ can be written as \bea\Phi_\chi = \int_{T_i = 0}^\infty d\Phi_\chi(T_i) \nonumber \eea
 
 Thus, 
 \bea
 \left(\frac{d\Phi_\chi}{dT_\chi}\right)_i = \int_{ 0}^\infty dT_i ~\frac{d\Phi_\chi}{dT_i}~ \frac{1}{T_\chi^{max}(T_i)}~ \Theta\left[T_\chi^{max}(T_i) - T_\chi \right] \nonumber \\
 \label{Eq:diffTchi}
 \eea
 
   The heaviside step function $\Theta\left[T_\chi^{max}(T_i) - T_\chi \right]$ inside the integration $\int_0^{\infty} dT_i \cdots $ ensures that $T_\chi^{max} > T_\chi$, where $T_\chi$ is a random variable having a flat distribution. In that way $\Theta\left[T_\chi^{max}(T_i) - T_\chi \right]$ turns the above integral to a more meaningful one  $\int_{T_i^{min}}^{\infty} \cdots$. In other words, the $\Theta$ function ensures that $T_\chi$ cannot take any value independent of $T_i$ and $T_i$ has a minimum value to produce a particular $T_\chi$.
So finally Eq.(\ref{Eq:diffTchi}) can be rewritten as 
 
 \bea
 \left(\frac{d\Phi_\chi}{dT_\chi}\right)_i = \int_{T_i^{min}}^\infty dT_i ~\frac{d\Phi_\chi}{dT_i}~ \frac{1}{T_\chi^{max}(T_i)}
 \eea
 
 Putting everything together, we get the following expression for the $i$-boosted DM flux from Eq.(\ref{Eq:diffTchi})
 \bea
 \left(\frac{d\Phi_\chi}{dT_\chi}\right)_i = D_{eff} \times \frac{\rho_\chi^{\rm{local}}}{m_\chi}~\sigma_{\chi e}~  \int^{\infty}_{T_i^{min}} dT_i ~ \frac{d\Phi_i}{dT_i}~ \frac{1}{T_\chi^{max}(T_i)} \nonumber \\
 \label{Eq:DMflux-wrt-T}
 \eea

 Using Eq.~(\ref{Eq:DMflux-wrt-T}), we obtain the BDM fluxes and show them Fig.~\ref{Fig:bdm_flux}. The bumps in the right panel of Fig.~\ref{Fig:bdm_flux} are due to the behaviour of DSNB spectrum. {We also notice that the flux of boosted sub-GeV DM peaks at energies greater than a MeV due to DSNB. Thus, DSNB contribution could be important in the context of Super-Kamiokande (where recoil energies are of order of a MeV's)}

\begin{figure*}[t]
\centering
\includegraphics[width=0.45\textwidth]{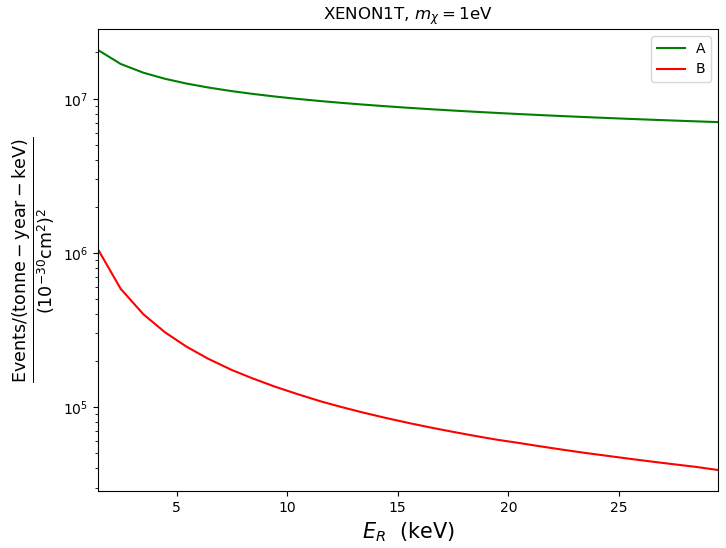}\hfill
\includegraphics[width=0.45\textwidth]{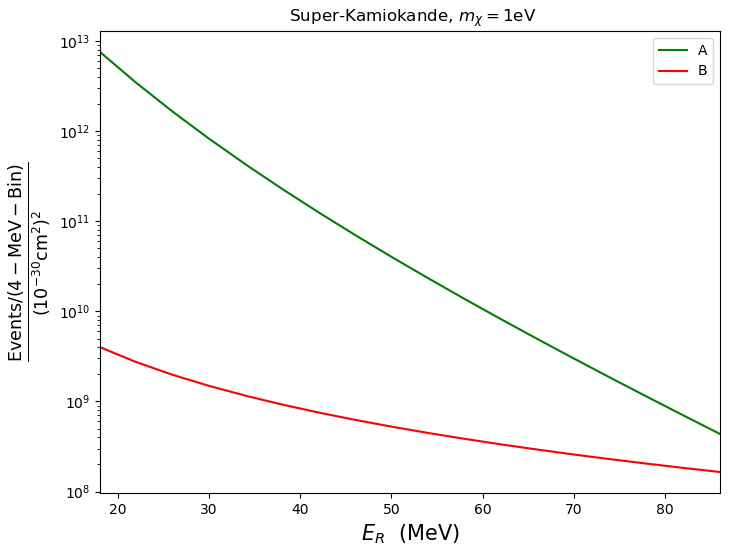} \\[1mm]
\includegraphics[width=0.45\textwidth]{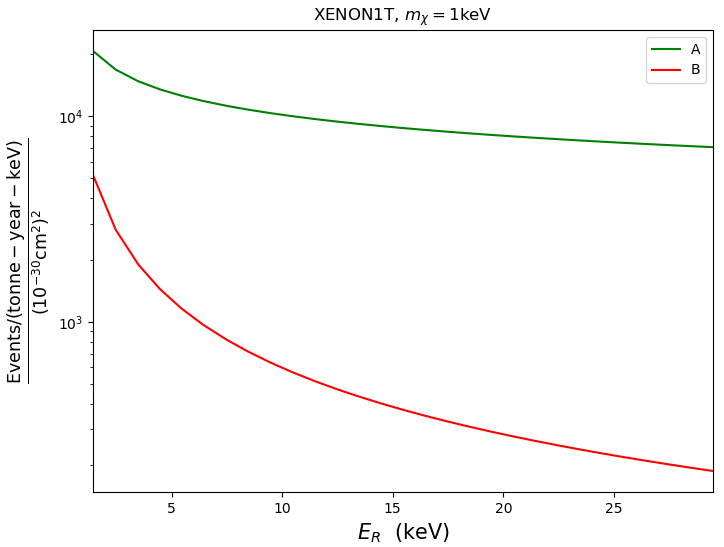}\hfill
\includegraphics[width=0.45\textwidth]{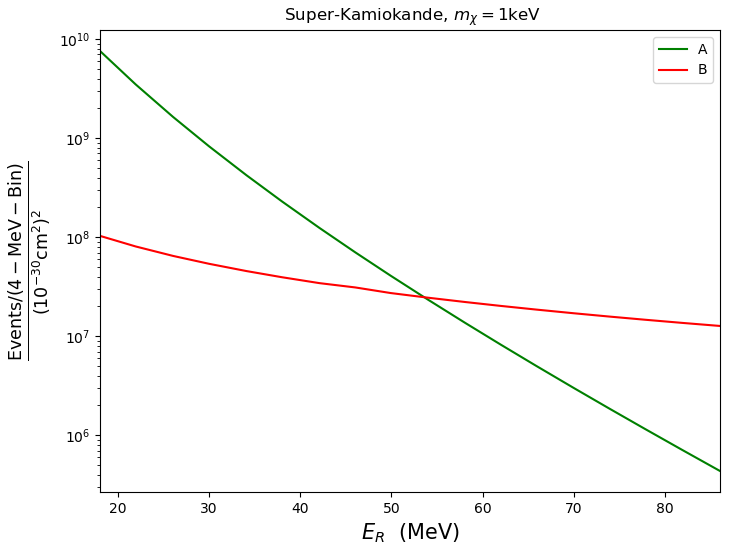} \\[1mm]
\includegraphics[width=0.45\textwidth]{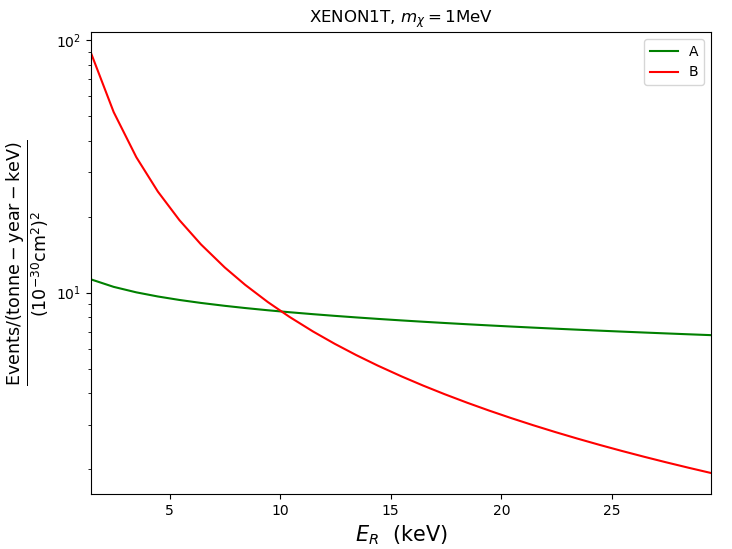}\hfill
\includegraphics[width=0.45\textwidth]{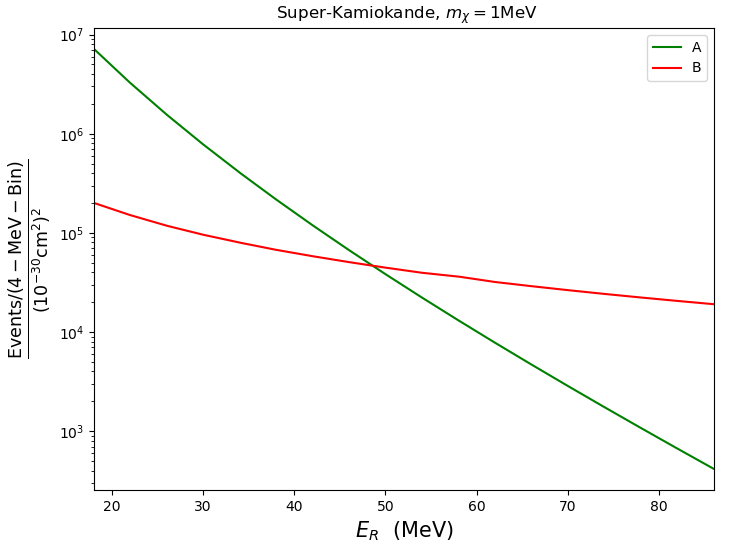}
\caption{Variations of A and B as a function of the recoil energy for $m_\chi = 1~\rm{eV}$ (top), $m_\chi = 1~\rm{keV}$ (middle), and $m_\chi = 1~\rm{MeV}$ (bottom). 
The left panel corresponds to the XENON1T experiment and the right panel corresponds to the Super-Kamiokande experiment.}
\label{Fig:fixed_mass}
\end{figure*}

\begin{figure*}[t]
\centering
\includegraphics[width=0.49\textwidth]{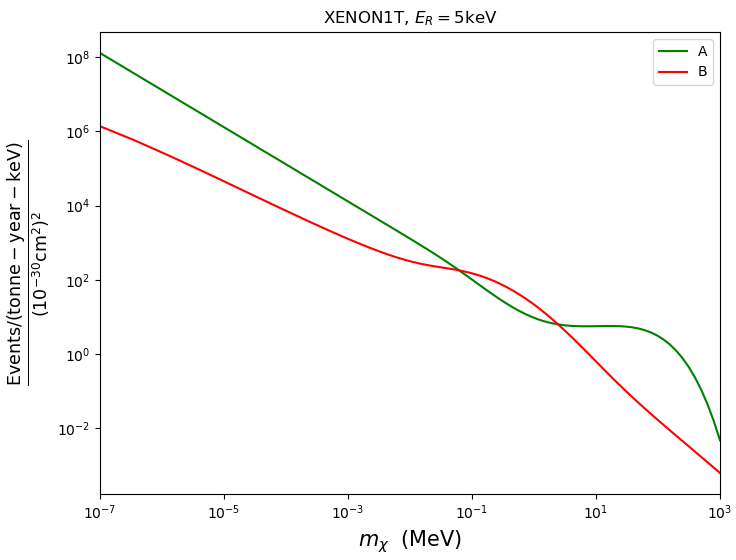}\hfill
\includegraphics[width=0.49\textwidth]{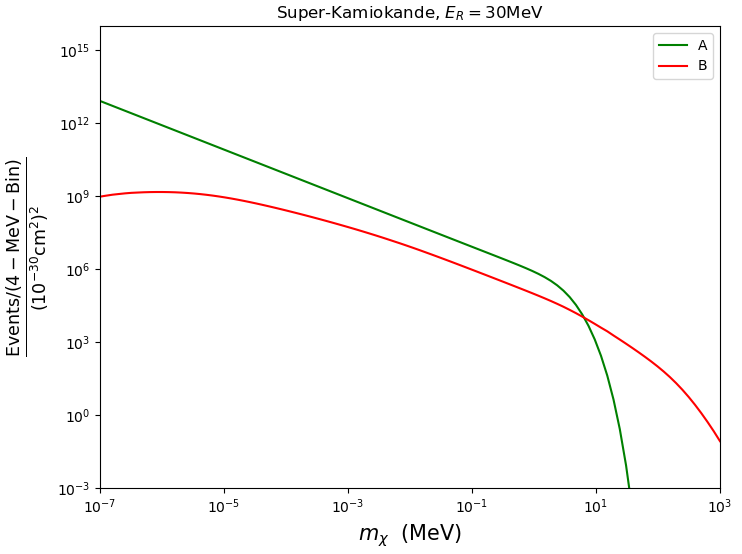}\\[1mm]
\includegraphics[width=0.49\textwidth]{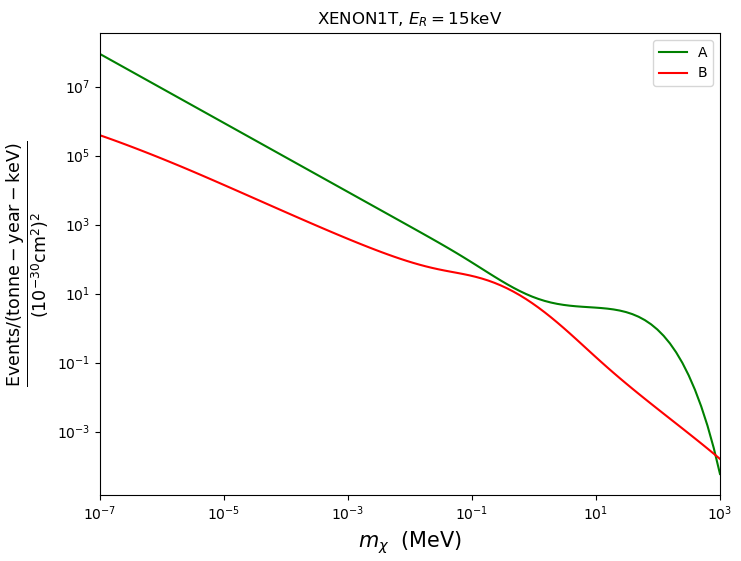}\hfill
\includegraphics[width=0.49\textwidth]{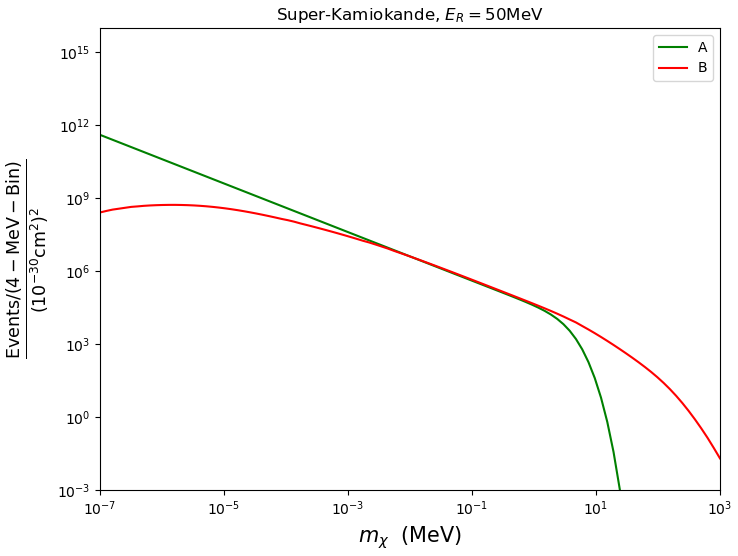}\\[1mm]
\includegraphics[width=0.49\textwidth]{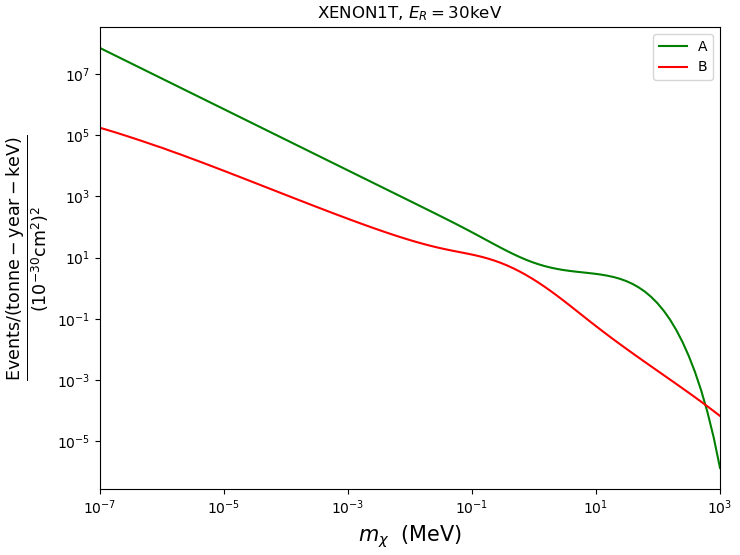}\hfill
\includegraphics[width=0.49\textwidth]{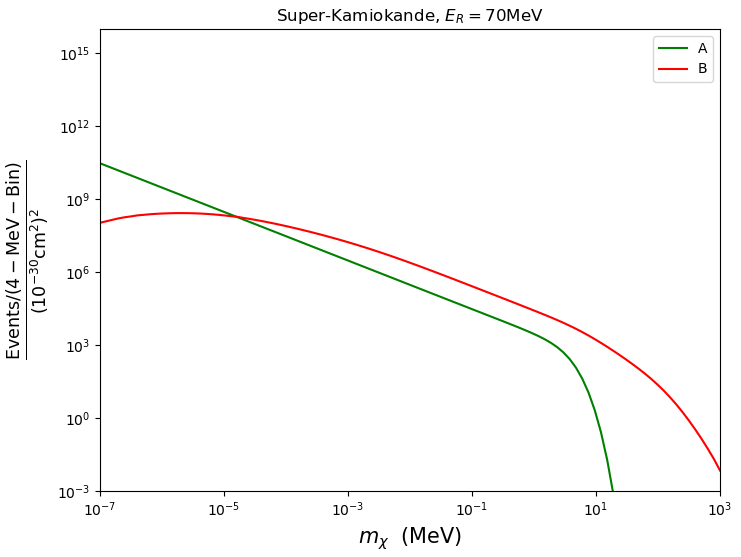}
\caption{Variation of A and B for XENON1T(left) and Super-Kamiokande(right) for different recoil energies. }
\label{Fig:fixed_ER}
\end{figure*}
%

\section{Rate equation}
\label{Sec:rate}

 Using the BDM flux obtained in Sec.~\ref{Sec:bdm_flux}, Eq.~(\ref{Eq:DMflux-wrt-T}), we get the differential recoil rate as follows
 
\bea
 \frac{dR}{dE_R} = \aleph\,\sigma_{\chi e} \int_{T_{\chi}^{min}(E_R)}^\infty dT_\chi \sum_{i=e,\nu} \left(\frac{d\Phi_\chi}{dT_\chi}\right)_i \frac{1}{E_R^{max}(T_\chi)} \nonumber \\
 \label{Eq:recoil}
 \eea  
where, $E_R^{max}(T_\chi)$ and $T_{\chi}^{min}(E_R)$ are obtained using Eqs.~(\ref{Eq:T-chi-max}) and (\ref{Eq:T-nu-min}) with the simple replacements $ i \to \chi$ and $\chi \to e$, respectively. As different experiments provide the event rate $R$ in different units, we incorporate this fact in a single expression (Eq.~\ref{Eq:recoil}) via parameter $\aleph$. Thus, the recoil spectrum for XENON1T is obtained by taking $\aleph\,=\,{Z_{Xe}}/{m_{Xe}}$, where $Z_{Xe}$ is  atomic number of Xenon and $m_{Xe}$ is the mass of a single Xenon atom. In case of the SK-I data, this factor corresponds to the total number of electrons, $\aleph\,=\,7.5 \times 10^{33}$~\cite{Super-Kamiokande:2011lwo,Ema:2018bih, Cappiello:2019qsw}.

Following Eq.(\ref{Eq:recoil}) we expand the recoil spectra for XENON1T and Super-Kamiokande to obtain
  \bea
 \frac{dR}{dE_R} = A \sigma_{\chi e} \sigma_{\chi \nu} + B \sigma_{\chi e}^2 
 \label{Eq:rate_expansion}
 \eea
 with
 \bea
 A = \aleph~D^\nu_eff~\frac{\rho_\chi}{m_\chi} \int\displaylimits_{T_{\chi}^{min}(E_R)}^\infty  \frac{dT_\chi}{E_R^{max}(T_\chi)} \int\displaylimits^{\infty}_{T_\nu^{min}}  \frac{dT_\nu}{T_\chi^{max}(T_\nu)}~ \frac{d\Phi_\nu}{dT_\nu} \nonumber \\
 \eea 
 and
 \bea
 B = \aleph~D^e_eff~\frac{\rho_\chi}{m_\chi} \int\displaylimits_{T_{\chi}^{min}(E_R)}^\infty  \frac{dT_\chi}{E_R^{max}(T_\chi)} \int\displaylimits^{\infty}_{T_e^{min}}  \frac{dT_e}{T_\chi^{max}(T_e)}~ \frac{d\Phi_e}{dT_e} \nonumber \\
 \eea 
where $A$ and $B$ are functions of DM mass and electron recoil energy. In Fig.~\ref{Fig:fixed_mass}, the coefficients A and B of Eq.~\ref{Eq:rate_expansion} have been plotted as a function of the recoil energy for three different DM masses, $m_\chi = 1~\rm{keV}, 1~\rm{MeV}, \text{and} \, \, 1~\rm{GeV}$ for both the experiments, XENON1T (left panel) and Super-Kamiokande (right panel). In Fig.~\ref{Fig:fixed_ER}, on the other hand, variations in A and B are shown as a function of the DM mass
for three different recoil energies. As it can be noticed from Fig.~\ref{Fig:fixed_mass}, for sub-MeV DM mass regime, $A$ dominates over $B$. Thus, the right side of Eq.~\ref{Eq:rate_expansion} gets most of the contribution from the $A$ term. This implies, for that mass range, the boost due to the scattering with DSNB neutrinos play the dominant role. This will also be clear in the next section.

\section{$\chi^2$ analysis} 

 To find the best-fit point and the exclusion region, we perform a $\chi^2$ analysis using the following definitions
 \bea
 \chi^2 &=& \sum_{i} \frac{(O_i - E_i)^2}{E_i + \left(\sigma_i^2\right)_{\rm{data}}} \\
 \Delta \chi^2 &=& \chi^2(BDM+B_0)-\chi^2(B_0~\rm{only})
 \eea
 
 where, $O_i$ are the observed number of events, $E_i$ are the expected number of events and $(\sigma_i)_{\rm{data}}$ is uncertainty in the measured data. For the $(BDM+B_0)$ case, to calculate the $E_i$ values, we sum the BDM signal and the background $B_0$ for each energy bin. 
 
\subsection*{Xenon1T}
\begin{figure}
\centering
\includegraphics[width=0.49\textwidth]{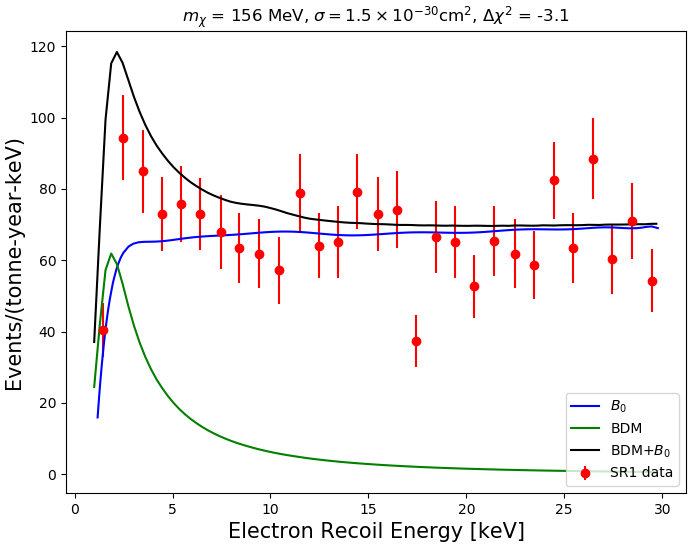}
\caption{Scientific Run 1 (SR1) data (in red) along with the estimated background, $B_0$ (in blue) reported by the XENON collaboration~\cite{XENON:2020rca}.
An example spectrum corresponding to the  best-fit point $(m_\chi, \sigma)=(156~\rm{MeV}, \, 1.5\times 10^{-30}~\rm{cm^2})$ is also shown (green and black) for DM boosted by CRe and DSNB neutrinos both.}
\protect\label{Fig:xenon_data}
\end{figure}

The XENON1T experiment is operated using a dual-phase liquid-xenon time projection chamber. The detector is capable of producing both prompt scintillation (S1) and delayed electroluminescence (S2) signals. The S2/S1 ratio is further used to distinguish electronic recoils from nuclear recoils. The XENON collaboration, last year, reported a $3.5\sigma$ excess of events in the electron recoil range of $1~\rm{keV} < E_R < 7~\rm{keV}$ with 285 events over the backgrounds of $232 \pm 15$ events. This data was acquired in total 277 days of live-time, which is referred to as Science Run 1 (SR1) \cite{XENON:2018voc,XENON:2020rca}. We use this data for the $\chi^2$ analysis to obtain the best-fit point as well as the exclusion region in ($m_\chi, \sigma$) plane. In Fig.~\ref{Fig:xenon_data}, we have shown the SR1 data (in red) along with the predicted background $B_0$ (in blue). The expected no.of events from boosted DM for the best-fit point along with the corresponding total number of events including the background are also shown. 

\begin{figure}[h!]
\centering
\includegraphics[width=0.49\textwidth]{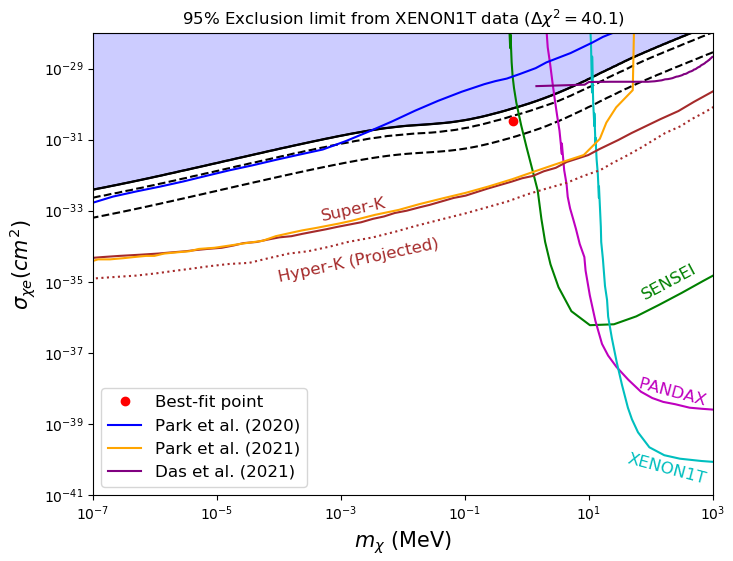}\hfill
\caption{Blue shaded exclusion region in the $(m_\chi\,,\sigma_{\chi e})$ plane derived from the XENON1T data at the $95\%$ confidence level for CRe boosted DM assuming $\sigma_{\chi \nu} = 0$. The region between the two dashed lines satisfy $\Delta \chi^2 < 0 $, with the best-fit point (0.6 MeV, $3.4\,\times\, 10^{-31}~\rm{cm^2}$) marked as a red point correspond to $\Delta \chi^2\,=\,-4.1 $. The constraints from other experiments on light cold DM such as SENSEI~\cite{SENSEI:2020dpa}, PANDAX II~\cite{PandaX-II:2021nsg}, XENON1T~\cite{XENON:2019gfn} are shown for comparison. We also give the results from  Ref.~\cite{Jho:2020sku}, Ref.~\cite{Jho:2021rmn} and Ref.~\cite{Das:2021lcr} for CRe , stellar neutrino and DSNB boosted DM derived from XENON1T, alongwith the results from Ref.~\cite{Cappiello:2019qsw} for CRe boosted DM derived from Super-K (also Hyper-K projection) data. 
}
\protect\label{Fig:chisqr_XENON1T_1}
\end{figure}

As it is evident from Fig.~\ref{Fig:xenon_data}, the expected background ($B_0$) can not explain the SR1 data. We estimated $\chi^2(B_0~\rm{only})$ to be $\sim 27.1$. Clearly,  if the BDM contribution explains SR1 data, $\Delta \chi^2$ must be less than $0$ commensurate with a better fit.   To derive the exclusion limit with the 95\% confidence, we demand $\Delta \chi^2 > 40.1$ that corresponds to 27 degrees of freedom. We present the obtained exclusion region in the  $m_\chi - \sigma_{\chi e}$ plane assuming $\sigma_{\chi \nu} = 0$ in Fig.~\ref{Fig:chisqr_XENON1T_1}. Our results agree very well with the one presented in \cite{Jho:2020sku} providing a good cross-check of our analysis. The best-fit point marked with red is given by $(m_\chi, \sigma)=(156~\rm{MeV}, \, 1.5\times 10^{-30}~\rm{cm^2})$ corresponding to $\Delta \chi^2 = -3.1 $. The constraints from various experiments searching for light cold DM such as SENSEI~\cite{SENSEI:2020dpa}, PANDAX II~\cite{PandaX-II:2021nsg}, XENON1T~\cite{XENON:2019gfn} are also shown for comparison. The results of Ref.~\cite{Jho:2020sku} and Ref.~\cite{Cappiello:2019qsw}, which derive limits on CRe boosted DM from XENON1T and Super-K data respectively, are also given. 

\begin{figure}[h!]
\centering
\includegraphics[width=0.49\textwidth]{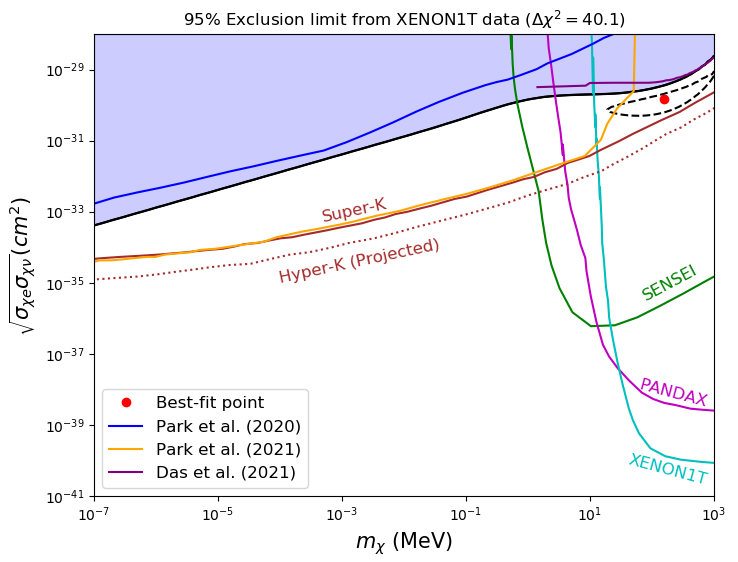}
\caption{Exclusion region in the $(m_\chi\,,\sqrt{\sigma_{\chi e}\sigma_{\chi\nu}})$ plane derived from the XENON1T data at the $95\%$ confidence level for DSNB boosted DM assuming $B = 0$ in Eq.~\ref{Eq:rate_expansion}. The region between the dashed contour satisfy $\Delta \chi^2 < 0 $, with the best-fit point (156 MeV, $1.5\times 10^{-30} {\rm cm^2}$) marked as a red point correspond to $\Delta \chi^2\,=\, -3.1$. 
 The constraints from other experiments on light cold DM such as SENSEI~\cite{SENSEI:2020dpa}, PANDAX II~\cite{PandaX-II:2021nsg}, XENON1T~\cite{XENON:2019gfn}, along with the constraints based on the results from Ref.~\cite{Jho:2020sku} for CRe BDM, Ref.~\cite{Jho:2021rmn} for stellar neutrino BDM and Ref.~\cite{Das:2021lcr} for DSNB BDM, derived from XENON1T data, are shown for comparison. We also give the results from Ref.~\cite{Cappiello:2019qsw} for CRe boosted DM derived from Super-K (also Hyper-K projection) data.  }
\protect\label{Fig:chisqr_XENON1T_12}
\end{figure}

Further, we explore the scenario where $\sigma_{\chi \nu,\chi e}\neq\,0$ and assume the flux of DSNB neutrinos is the sole contributor to the DM boost i.e., $B = 0$ in Eq.(\ref{Eq:rate_expansion}). To do this, we repeat the above analysis and obtain a bound on $\sqrt{\sigma_{\chi \nu} \sigma_{\chi e}}$ as a function of DM mass. The results are shown in Fig.~\ref{Fig:chisqr_XENON1T_12} and we compare these with the results of Ref.~\cite{Das:2021lcr} which is analogous to the considered case and of Ref.~\cite{Jho:2021rmn} in which stellar neutrinos instead of DSNB neutrinos, boost the DM. Note that our results are consistent with Ref.~\cite{Das:2021lcr} for large $m_\chi$ values. { Also note that XENON1T sets bound $\sigma_{\chi e}\leq10^{-40}{\rm cm}^2$ for $m_\chi\,>\,100\,{\rm MeV}$ whereas bounds from SN1987A on new particles that have feeble interactions with electrons in addition to stronger-than-weak interactions with neutrinos allows $\sigma_{\chi\nu}\leq 10^{-25}{\rm cm}^2(m_\chi/{\rm MeV})$~\cite{Boehm:2013jpa,Bertoni:2014mva}\footnote{We thank anonymous referee  for pointing out these bounds}. Therefore, the best-fit point obtained for $B = 0$ case is ruled out.}

\begin{figure}
\centering
\includegraphics[width=0.49\textwidth]{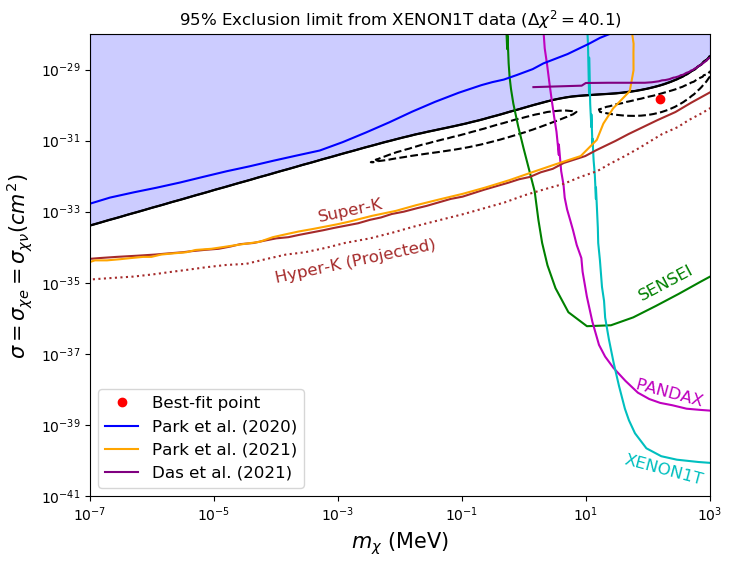}\hfill
\caption{Exclusion region in the $(m_\chi\,,\sigma_{\chi \nu} = \sigma_{\chi e})$ plane derived from the XENON1T data at the $95\%$ confidence level for DM boosted by both CRe and DSNB, corresponding to $A \neq 0$ and $B \neq 0$ in Eq.~(\ref{Eq:rate_expansion}). For all points inside the dashed line $\Delta \chi^2 < 0 $ and the best fit point (156 MeV, $1.5\times 10^{-30} {\rm cm^2}$) marked with red point correspond to $\Delta \chi^2\,=\,-3.1$.  The constraints from other experiments on light cold DM such as SENSEI~\cite{SENSEI:2020dpa}, PANDAX II~\cite{PandaX-II:2021nsg}, XENON1T~\cite{XENON:2019gfn}, along with the constraints based on the results from Ref.~\cite{Jho:2020sku} for CRe BDM, Ref.~\cite{Jho:2021rmn} for stellar neutrino BDM and Ref.~\cite{Das:2021lcr} for DSNB BDM, derived from XENON1T data, are shown for comparison. We also give the results from Ref.~\cite{Cappiello:2019qsw} for CRe boosted DM derived from Super-K (also Hyper-K projection) data. }
\protect\label{Fig:chisqr_XENON1T_2}
\end{figure}

As we mentioned earlier, the flux of DSNB neutrinos is comparable to the flux of CRe in that energy range which provides visible recoil energy due to light DM scattering off electron. Therefore, next, we consider contribution of both the fluxes. We assume $\sigma_{\chi \nu} = \sigma_{\chi e}$  and derive the exclusion region, presented in Fig.~\ref{Fig:chisqr_XENON1T_2}. 

Note that the exclusion region in Fig.~\ref{Fig:chisqr_XENON1T_1} and Fig.~\ref{Fig:chisqr_XENON1T_12} is derived from $A$ and $B$ term of Eq.~\ref{Eq:rate_expansion} respectively. In the current case, both terms are combined to set the exclusion limit. For lighter $m_\chi$, $A$ term wins over $B$, shown in Fig.~\ref{Fig:fixed_mass}. Thus, the numbers in Fig.~\ref{Fig:chisqr_XENON1T_12} and Fig.~\ref{Fig:chisqr_XENON1T_2} are similar.

\begin{figure}
\centering
\includegraphics[width=0.49\textwidth]{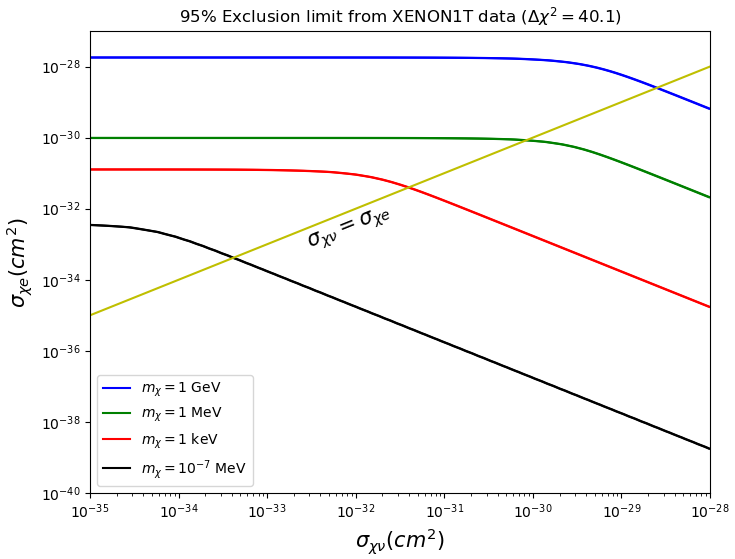}
\caption{Exclusion contours in $(\sigma_{\chi \nu}\,, \sigma_{\chi e})$ plane derived from the XENON1T data for DM boosted by both CRe and DSNB are shown for different values of $m_\chi$. The regions above the solid lines are excluded at the $95\%$ confidence level.  }
\protect\label{Fig:chisqr_XENON1T_22}
\end{figure}

We also present the exclusion contours in $\sigma_{\chi \nu}- \sigma_{\chi e} $ plane for different values of $m_\chi$ in Fig.~\ref{Fig:chisqr_XENON1T_22}. The region above the lines is excluded at the 95\% confidence level. In the regime $\sigma_{\chi \nu}\gg\sigma_{\chi e}$, the first term in Eq.~\ref{Eq:rate_expansion} dominates which implies $\sigma_{\chi e}\sigma_{\chi \nu}\,=\,{\rm constant}$ and thus is the nature of the curve.  For $\sigma_{\chi e}\gg\sigma_{\chi \nu}$, the second term in Eq.~\ref{Eq:rate_expansion} dominates and we obtain a contour independent of $\sigma_{\chi \nu}$. Also, area excluded increases with decreasing DM mass because lighter DM is boosted more and hence it is more constrained. 

\begin{figure}[h!]
\centering
\includegraphics[width=0.49\textwidth]{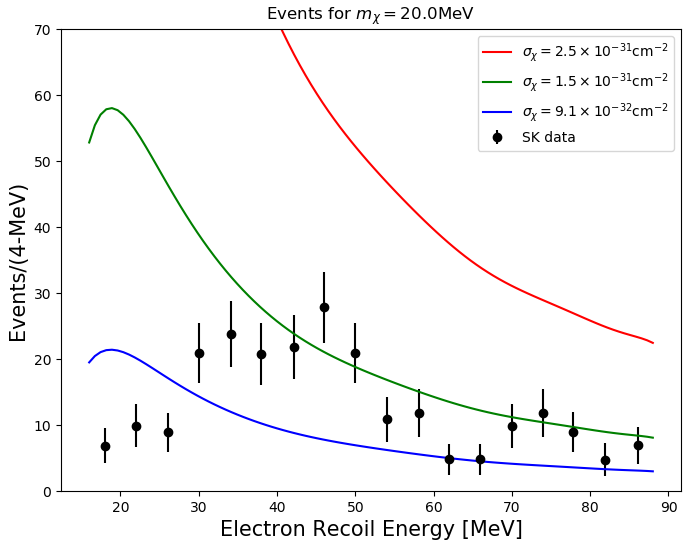}\hfill
\includegraphics[width=0.49\textwidth]{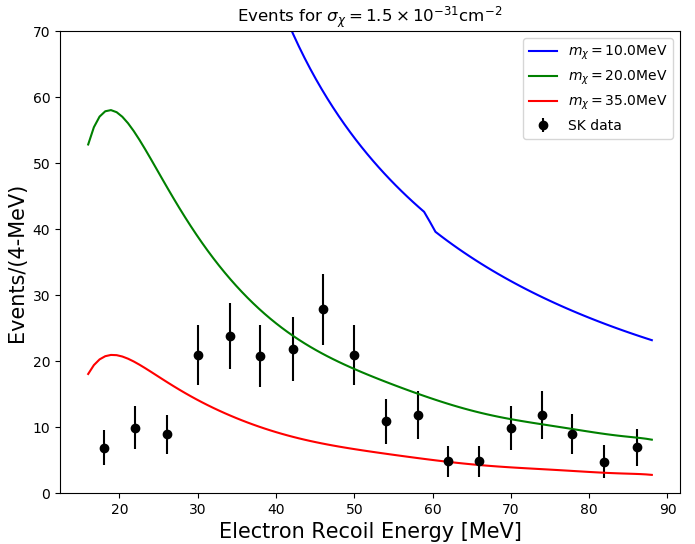}
\caption{Super-Kamiokande data~\cite{Super-Kamiokande:2011lwo} along with the predicted no. of events for DM boosted via CRe and DSNB neutrinos. In the upper panel, the no. of events are shown as a function of the electron recoil energy for different values of cross-section at fixed $m_\chi=20\,{\rm MeV}$. In the bottom panel,  the event rate as a function of the recoil energy is given for different values of $m_\chi$ at fixed $\sigma\,=\,1.5\times10^{-31}\,{\rm cm^2}$. The visible wiggles are present due to the detector efficiency.}
\protect\label{Fig:sk_data}
\end{figure}

\begin{figure}
\centering
\includegraphics[width=0.49\textwidth]{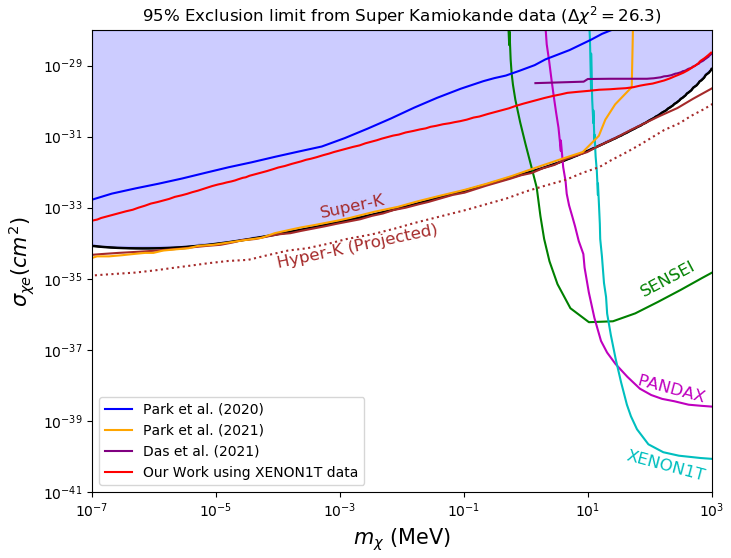}
\caption{Exclusion region in the $(m_\chi\,,\sigma_{\chi e})$ plane derived from the SK I data at the $95\%$ confidence level for CRe boosted DM assuming $\sigma_{\chi \nu} = 0$. The constraints from other experiments on light cold DM such as SENSEI~\cite{SENSEI:2020dpa}, PANDAX II~\cite{PandaX-II:2021nsg}, XENON1T~\cite{XENON:2019gfn}, along with the constraints based on the results from Ref.~\cite{Jho:2020sku} for CRe BDM, Ref.~\cite{Jho:2021rmn} for stellar neutrino BDM and Ref.~\cite{Das:2021lcr} for DSNB BDM, derived from XENON1T data, are shown for comparison. We also give the results from Ref.~\cite{Cappiello:2019qsw} for CRe boosted DM derived from Super-K (also Hyper-K projection) data. }
\protect\label{Fig:chisqr_sk_1}
\end{figure}

\begin{figure}
\centering
\includegraphics[width=0.49\textwidth]{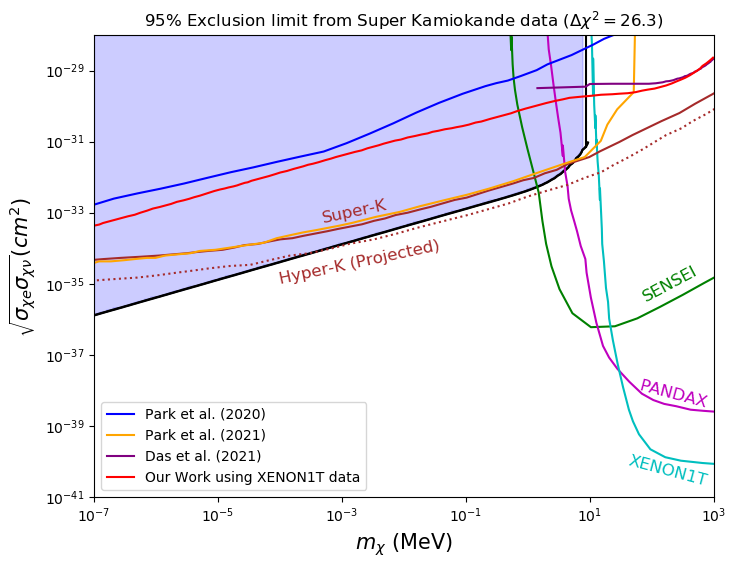}
\caption{Exclusion region in the $(m_\chi\,,\sqrt{\sigma_{\chi e}\sigma_{\chi\nu}})$ plane derived from the SK I data at the $95\%$ confidence level for DSNB boosted DM assuming $B = 0$ in Eq.~\ref{Eq:rate_expansion}.  The constraints from other experiments on light cold DM such as SENSEI~\cite{SENSEI:2020dpa}, PANDAX II~\cite{PandaX-II:2021nsg}, XENON1T~\cite{XENON:2019gfn}, along with the constraints based on the results from Ref.~\cite{Jho:2020sku} for CRe BDM, Ref.~\cite{Jho:2021rmn} for stellar neutrino BDM and Ref.~\cite{Das:2021lcr} for DSNB BDM, derived from XENON1T data, are shown for comparison. We also give the results from Ref.~\cite{Cappiello:2019qsw} for CRe boosted DM derived from Super-K (also Hyper-K projection) data. }
\protect\label{Fig:chisqr_sk_12}
\end{figure}

\begin{figure}
\centering
\includegraphics[width=0.49\textwidth]{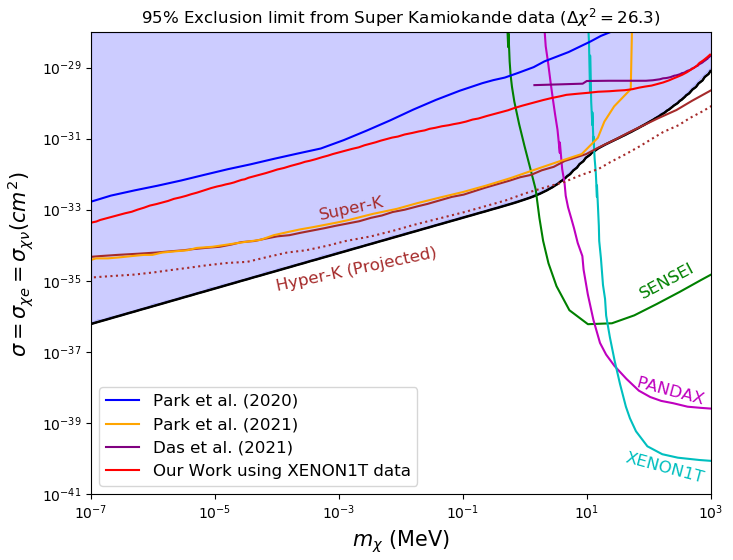}
\caption{Exclusion region in the $(m_\chi\,,\sigma_{\chi \nu} = \sigma_{\chi e})$ plane derived from the SK I data at the $95\%$ confidence level for DM boosted by both CRe and DSNB, corresponding to $A \neq 0$ and $B \neq 0$ in Eq.~(\ref{Eq:rate_expansion}). The constraints from other experiments on light cold DM such as SENSEI~\cite{SENSEI:2020dpa}, PANDAX II~\cite{PandaX-II:2021nsg}, XENON1T~\cite{XENON:2019gfn}, along with the constraints based on the results from Ref.~\cite{Jho:2020sku} for CRe BDM, Ref.~\cite{Jho:2021rmn} for stellar neutrino BDM and Ref.~\cite{Das:2021lcr} for DSNB BDM, derived from XENON1T data, are shown for comparison. We also give the results from Ref.~\cite{Cappiello:2019qsw} for CRe boosted DM derived from Super-K (also Hyper-K projection) data. }
\protect\label{Fig:chisqr_sk_2}
\end{figure}

\begin{figure}
\centering
\includegraphics[width=0.49\textwidth]{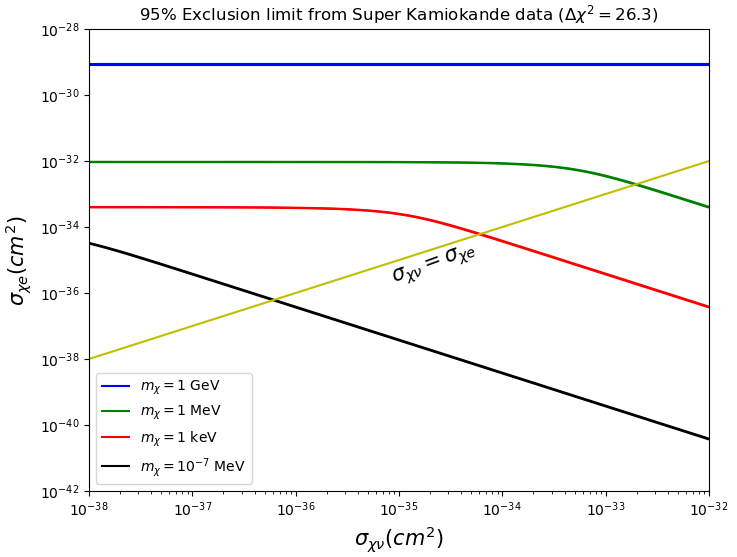}
\caption{Exclusion contours in $(\sigma_{\chi \nu}\,, \sigma_{\chi e})$ plane derived from the SK I data for DM boosted by both CRe and DSNB are shown for different values of $m_\chi$. The regions above the solid lines are excluded at the $95\%$ confidence level.}
\protect\label{Fig:chisqr_sk_22}
\end{figure}

\subsection*{Super Kamiokande}
Super-K is a 50 kiloton water Cherenkov detector build at the Kamioka mine in Japan. The data used for this analysis is referred to as the SK-I data, which was taken for total 1497 days of live-time. The detector directly looks for the DSNB events via inverse beta decay $\bar{\nu_e} + p \to n + e^+ $. In the present work, we assume that the observed events are consistent with the background and hence the signal due to DM should be consistent with the data within the uncertainity. Since an estimate of the background is not found in the literature (to our knowledge), we take $\chi^2(B_0~\rm{only}) = 0$ for SK-I data. In Fig.~\ref{Fig:sk_data}, we have shown the SK-I data  along with the predicted no. of events for CRe + DSNB boosted DM for different values of $m_\chi$. Also shown are BDM-induced recoil spectra for different values of $\sigma_{\chi e}$ for $m_\chi=20$ MeV.  

We follow the $\chi^2$ analysis similar to XENON1T case to obtain the constraints on the $(m_\chi , \sigma)$ parameter space. We, first, consider the scenario when $\sigma_{\chi \nu} = 0$ and present the results in Fig.~\ref{Fig:chisqr_sk_1}. Here, the excluded region satisfies $\Delta \chi^2 > 26.3$ which corresponds to $95\%$ exclusion limit for 16 degrees of freedom. As it is evident from Fig.~\ref{Fig:chisqr_sk_1}, our  results agree very well with that of Ref.~\cite{Cappiello:2019qsw}. 

In Fig.~\ref{Fig:chisqr_sk_12}, we show the exclusion region assuming $B = 0$ in Eq.(\ref{Eq:rate_expansion}) or the flux of DSNB neutrinos is the only contributor to the DM boost. To best of our knowledge, there is no previous study where the exclusion limits on $\sigma_{\chi \nu}$ are derived using SK-I data. Also, these bounds are stronger than those obtained in earlier studies for different detectors. 

It should be noticed that the exclusion region extends beyond 1 GeV in Fig.~\ref{Fig:chisqr_sk_1} which correspond to the CRe boosted DM scenario whereas it is limited to 10 MeV in the  case of DSNB boosted DM (Fig.~\ref{Fig:chisqr_sk_12}). This owes to the fact that DSNB flux decline very rapidly for $T_\nu\,>\,50\,{\rm MeV}$ whereas cosmic electron spectra is comparatively significant at larger kinetic energy. 

Next, we consider the DSNB + CRe flux which boost the DM and derive the most general exclusion limits. Assuming $\sigma_{\chi \nu} = \sigma_{\chi e}$, we present the exclusion regions and contours in Fig.~\ref{Fig:chisqr_sk_2} and Fig.~\ref{Fig:chisqr_sk_22} respectively. Note that numbers obtained in this case are comparable to that of Fig.~\ref{Fig:chisqr_sk_1} for larger masses whereas are similar to DSNB boosted DM case (Fig.~\ref{Fig:chisqr_sk_12}) for $m_\chi\,<\,10\,{\rm MeV}$. To understand this, we follow Fig.~\ref{Fig:fixed_mass} for SK I data. We observe that $A$ term of Eq.~\ref{Eq:rate_expansion} dominates over $B$ term for DM of mass less than a few MeV, for $E_R\,\leq\,50\,{\rm MeV}$. Not only does it dominate in certain energy range, its magnitude is very large. Therefore, bounds obtained in Fig.~\ref{Fig:chisqr_sk_2} for $m_\chi < 10\,{\rm MeV}$ replicate the bounds in Fig.~\ref{Fig:chisqr_sk_12}. 

\section{Conclusions}
In this work, we implemented the idea of boosted DM to set exclusion limits on a combination of DM-electron and DM-neutrino cross-sections for low-mass DM. To register events in the detectors of such DM particle interacting with neutrinos, obviously, we need to assume non-zero interaction strength between DM and the  electrons. Therefore we can not ignore the boost due to scattering of DM with CRe while constraining DM-neutrino interactions. Furthermore, we also noted that the flux of the DSNB boosted-DM and the CRe boosted-DM are comparable for light DM in the energy range relevent for XENON1T and Super-Kamiokande. We perform a $\chi^2$ analysis to obtain novel limits in the $(m_\chi,\sigma_{\chi e/\chi\nu})$ plane using XENON1T (a low energy recoil experiment) and Super-K (a high energy recoil experiment) data. We also find the best-fit points explaining the reported excess events by XENON1T collaboration. We systematically study the following cases: (i) cosmic electron boosted DM where only DM-electron interaction is considered, (ii) a scenario where DM particles get boosted only due to their interactions with the neutrinos, and (iii) DM is boosted due to interactions with cosmic electron as well as DSNB. To our knowledge, we are the first to use Super-K data to derive bounds on $\sigma_{\chi \nu}$. We found that Super-Kamiokande, in fact, sets the strongest bound on $\sigma_{\chi \nu}$ for $m_\chi\,<\,10\,{\rm MeV}$, as shown in Fig.~\ref{Fig:chisqr_sk_12} and Fig.~\ref{Fig:chisqr_sk_2}. { It should be noted that while the DSNB neutrino contribution dominate the Super-K limits, the stellar neutrino contribution to boosted DM (given in Ref.~\cite{Jho:2021rmn}) flux can lead to a non-negligible improvement in bounds obtained via XENON1T since it is a low-energy recoil experiment and stellar neutrino flux peaks at energies in keV range.  However, we leave a dedicated analysis on CRe+DSNB+Stellar neutrino, a combined effect, for future work.}

We believe that large part of the parameter space for light boosted DM could also be probed with other low-energy recoil experiments like SENSEI~\cite{SENSEI:2020dpa}, CRESST-II~\cite{CRESST:2015txj}, PandaX~\cite{PandaX-II:2021nsg} etc., the experiments which usually probe cold DM. Similarly, it would be interesting to see how other neutrino experiments/detectors like  Borexino~\cite{Bellini:2011rx}, DUNE~\cite{DUNE:2015lol}, JUNO~\cite{JUNO:2015zny}, Hyper-Kamiokande~\cite{Hyper-Kamiokande:2016srs}, MiniBooNE~\cite{MiniBooNEDM:2018cxm} could be utilized to rule out the models of boosted DM. 

\section*{acknowledgments}
DG and DS acknowledge support through the Ramanujan Fellowship and MATRICS Grant of the Department of Science and Technology, Government of India.  DS has in-part received funding from the European Union’s Horizon 2020 research and innovation programme under grant agreement No 101002846, ERC
CoG “CosmoChart``. DG and AG would like to thank Arun Thalapillil for discussions at the early stage of the work. A.G. acknowledges support from the National Research Foundation of Korea (NRF-2019R1C1C1005073). A.G. also wants to thank Jong-Chul Park for useful discussions.

\bibliographystyle{apsrev4-1}
\bibliography{bdm_ref}

\end{document}